\documentstyle[12pt]{article}
\def\mathbf{\vec}

\def\ca{\c{c}\~{a}}

\begin{document}

\centerline {\LARGE Effective Chiral Meson Lagrangian}
\vspace{.3cm}
\centerline {\LARGE For The Extended Nambu -- Jona-Lasinio Model}
\vspace{1cm}
\centerline {\large Alexander A. Osipov\footnote{On leave from the 
            Laboratory of Nuclear Problems, JINR, Dubna, Russia}, 
            Brigitte Hiller}
\vspace{.5cm}
\centerline {\it Centro de F\'{\i}sica Te\'{o}rica, Departamento de
             F\'{\i}sica}
\centerline {\it da Universidade de Coimbra, 3004-516 Coimbra, Portugal}
\vspace{1cm}

\begin{abstract}
We present a derivation of the low-energy effective meson Lagrangian of the
extended Nambu -- Jona-Lasinio (ENJL) model. The case with linear realization
of broken $SU(2)\times SU(2)$ chiral symmetry is considered. There are two 
crucial points why this revision is needed. Firstly it is the explicit chiral 
symmetry breaking effect. On the basis of symmetry arguments we show that 
relevant contributions related with the current quark mass terms are absent
from the effective Lagrangians derived so far in the literature. Secondly we 
suggest a chiral covariant way to avoid non-diagonal terms responsible for the 
pseudoscalar -- axial-vector mixing from the effective meson Lagrangian. In
the framework of the linear approach this diagonalization has not been done 
correctly. We discuss as well the $SU(2)\times SU(2)/SU(2)$ coset space 
parametrization for the revised Lagrangian (nonlinear ansatz). Our Lagrangian 
differs in an essential way from those that have been derived till now on the 
basis of both linear and nonlinear realizations of chiral symmetry.
\end{abstract}

%%%%%%%%%%%%%%%%%%%%%%%%%%%%%%%%%%%%%%%%%%%%%%%%%%%%%%%%%%%%%%%%%%%%%%%%%%%%%%%
\vspace{4.0mm}
\noindent
PACS number(s): 12.39.Fe, 11.30.Rd.
\vspace{4.0mm}
%%%%%%%%%%%%%%%%%%%%%%%%%%%%%%%%%%%%%%%%%%%%%%%%%%%%%%%%%%%%%%%%%%%%%%%%%%%%%%%

\newpage
%%%%%%%%%%%%%%%%%%%%%%%%%%%%%%%%%%%%%%%%%%%%%%%%%%%%%%%%%%%%%%%%%%%%%%%%%%%%%%%
\section{Introduction}
%%%%%%%%%%%%%%%%%%%%%%%%%%%%%%%%%%%%%%%%%%%%%%%%%%%%%%%%%%%%%%%%%%%%%%%%%%%%%%%

The Nambu -- Jona-Lasinio (NJL) model \cite{Nambu:1961} is useful because it 
allows to derive the effective meson Lagrangian from a more fundamental, i.e. 
microscopic, theory of quarks. The effective four-fermion interactions
of the NJL-like models represent ``certain approximations" to QCD. From the
theoretical point of view, however, it is still not clear in which way these
four-quark interactions arise in QCD. In the case of two flavours one of the 
possible mechanisms might be the quarks' interaction via the zero modes of 
instantons \cite{Diakonov:1995}, the so-called 't Hooft interactions. 
Nevertheless there are a lot of investigations directed to the low-energy 
hadron phenomenology following from NJL-like Lagrangians 
\cite{Ebert:1983}-\cite{Bernard:1996}. The reasons are clear and well-known. 
These approximations are much easier to handle than QCD. They provide us with 
an unique way of constructing effective meson Lagrangians including vector and
axial-vector mesons. They incorporate most of the short-distance relations
which follow from QCD. In addition the NJL-like models are a good playground 
from the mathematical point of view. Starting from the basic quark Lagrangian
one can develop both the techniques of the linear 
\cite{Ebert:1983,Volkov:1984,Ebert:1986} and nonlinear \cite{Bijnens:1993} 
realizations of chiral symmetry. Both parametrizations for the chiral fields 
must lead to the same predictions and are equivalent on the mass-shell. The 
integration over the quark fields in the generating functional yields the 
determinant of the Dirac operator $D$ in the presence of bosonic fields. Its 
evaluation must conform with the chiral covariant formulation of quantum field 
theory. The difficulties encountered in the realization of this idea are 
reviewed in \cite{Ball:1989}. 

In order to calculate the effective action and study spontaneous breakdown of 
global chiral symmetry it is important to employ a method of calculation which 
preserves the symmetry explicitly. It is known that the Schwinger proper-time 
representation \cite{Schwinger:1951,DeWitt:1965} for $\ln |\det D|$ in terms of
the modulus of the quark determinant and the following long wavelength 
expansion of its heat kernel fulfills this requirement. This technique is 
especially good to describe the low-energy regime of QCD \cite{Gasser:1984}.
However, in the presence of the explicit chiral symmetry breaking term in the
Lagrangian, the standard definition of $\ln |\det D|$ in terms of a proper-time 
integral 
\begin{equation}
\label{logdet}
   \ln |\det D|=-\frac{1}{2}\int^\infty_0\frac{dT}{T}\rho (T,\Lambda^2)
                 \mbox{Tr}\left(e^{-TD^\dagger D}\right)
\end{equation}
modifies the explicit chiral symmetry breaking pattern of the original quark 
Lagrangian and needs to be corrected in order to lead to the fermion 
determinant whose transformation properties exactly comply with the symmetry 
content of the basic Lagrangian \cite{Osipov:2000}. The necessary modifications 
can be done by adding a functional in the collective fields and their 
derivatives to the definition of the real part of the fermion determinant, i.e.
we define that
\begin{equation}
\label{newlndet}
   \mbox{Re}(\ln\det D)=\ln |\det D|+P.
\end{equation}
In the limit $\hat{m}=0$, where $\hat{m}$ is a current quark mass, $P=0$ and 
the old result (\ref{logdet}) emerges as a part of our definition. This 
strategy reminds Gasser and Leutwyler's correcting procedure which they used 
however for a different purpose, namely to restore the standard result 
(\ref{logdet}) for the real part of the fermion determinant defined by the heat
kernel $\mbox{Tr}[\mbox{exp}(-T\bar{D}^2)]$, especially chosen to include 
anomalies \cite{Gasser:1984}. Both of these procedures are aimed at the 
subtraction of inessential contributions inherent to the starting definitions of
$\det D$. These contributions are inessential in the sense that they change the 
content of the theory, what should not be. The procedures in \cite{Gasser:1984}
and ours differ however through the way of fixing the form of the functional 
$P$, because the origin of these contributions is different. In the case under 
consideration the functional $P$ must be chosen in such a manner that the real 
part of the effective Lagrangian for the bosonized ENJL model 
${\cal L}_{\mbox{eff}}$ will have the same transformation laws as the basic 
quark Lagrangian ${\cal L}$. In addition it should not change the 
``gap"-equation, i.e. the Schwinger -- Dyson
equation which defines the vacuum state of the model. These requirements 
together completely fix the freedom inherent to the definition of this 
functional. Let us stress that in our case $P$ cannot be fixed by the 
requirement that the determinant remains unchanged when axial-vector and 
pseudoscalar fields are switched off, like, for instance, in 
\cite{Gasser:1984}. As a consequence $P$ contributes to the effective potential
of the NJL model at every step of the heat kernel expansion. We have $P$ being 
a functional as opposed to a polynomial in \cite{Gasser:1984}. This is a 
general feature related to the non-renormalizability of the NJL model. Using 
formula (\ref{newlndet}) together with the way we propose to fix $P$, one can 
systematically take into account the effect of explicit chiral symmetry 
breaking in the ENJL model. To show this is one of the reasons for this paper. 
The correct description of explicit chiral symmetry breaking is evidently 
necessary in order to obtain realistic mass formulae and meson dynamics. We 
derive these expressions here and show that they are different (already in the 
leading current quark mass dependent part) from the results known in the 
literature.

The second reason for this work is related to the problem of the 
pseudoscalar -- axial-vector
mixing in the ENJL model. For some reason this diagonalization has never been 
done correctly in the framework of the linear realization of chiral symmetry,  
as it has been already indicated in \cite{Ecker:1989}. The usual procedure 
recurs to a linearized transformation 
\begin{equation}
\label{wrong}
  a_\mu\rightarrow a_\mu +c\partial_\mu\pi
\end{equation} 
which ruins the chiral transformation properties of the field $a_\mu$ and gives
rise to all sorts of apparent symmetry breaking. For example, it leads to the
$\rho\pi\pi$ coupling of the form $\rho_\mu [\pi ,\partial_\mu\pi ]$, which
breaks chiral symmetry. Here we suggest instead a covariant way to avoid 
non-diagonal terms responsible for the pseudoscalar -- axial-vector mixing in 
the effective meson Lagrangian. The covariant redefinition of the axial-vector 
field cannot be done without a corresponding change in its chiral partner, i.e.
the vector field. This is a direct consequence of the linear realization of 
chiral symmetry. We have found two bilinear combinations of scalar and 
pseudoscalar fields which transform like axial-vector and vector fields and 
are chiral partners at the same time. We also show that our procedure, if one 
rewrites it in the new coset space variables corresponding to the nonlinear 
representation of the chiral group, is identical to the one already known from 
\cite{Bijnens:1993} or \cite{Gas:1969}.
 
As a result we get the effective meson Lagrangian of the ENJL model in a form
which includes only the first three $(a_0, a_1, a_2)$ Seeley -- DeWitt 
coefficients in the asymptotic expansion for the heat kernel. We restrict
to this approximation, the extension is straightforward. Because of the 
aforementioned reasons we obtain a new revised Lagrangian which obeys all 
symmetry requirements of the model for the case of linear realization of broken
$SU(2)\times SU(2)$ chiral symmetry. We derive as well the $SU(2)\times 
SU(2)/SU(2)$ coset space parametrization for the revised Lagrangian. For that 
purpose the Lagrange multiplier method is used to eliminate the scalar field 
from the generating functional, thus arriving to the nonlinear version of the 
model.

The plan of the paper is the following: In Sec.2 we discuss the Lagrangian of 
ENJL model and show that chiral $SU(2)\times SU(2)$ transformations of quark 
fields dictate the transformation laws of the auxiliary bosonic fields. These 
collective variables are necessary to rearrange the four-quark Lagrangian of 
the ENJL model in an equivalent Lagrangian which is only quadratic in the 
quark fields. In Sec.3 we show how to define the fermion determinant for the 
case in which explicit symmetry breaking takes place. We calculate the first
three contributions in the asymptotic expansion of the heat kernel in full
detail. We derive the corresponding correcting polynomial from the functional 
$P$  and show that it is 
completely fixed by the symmetry breaking pattern of the basic quark Lagrangian
and the requirement that $P$ should not change the ``gap"- equation. The 
effective meson Lagrangian ${\cal L}_{\mbox{eff}}$ is obtained at the end of 
this section. In Sec.4 we introduce the new variables for vector and 
axial-vector fields in order to avoid the pseudoscalar -- axial-vector mixing 
term from  ${\cal L}_{\mbox{eff}}$. We use chiral covariant combinations for 
this replacement. We discuss the field renormalizations needed to define the 
physical meson states and the meson mass spectrum. The transition to the 
nonlinear version is done in Sec.5. The concluding remarks are given in Sec.6. 
Finally we show in the Appendix that the replacements of variables done in 
Sec.4 for the spin one mesons are completely equivalent to the replacement 
which has been already used in the literature in the context of the non-linear 
parametrization in the chiral group space.

%%%%%%%%%%%%%%%%%%%%%%%%%%%%%%%%%%%%%%%%%%%%%%%%%%%%%%%%%%%%%%%%%%%%%%%%%%%%%%%
\section{Lagrangian and its symmetries}
%%%%%%%%%%%%%%%%%%%%%%%%%%%%%%%%%%%%%%%%%%%%%%%%%%%%%%%%%%%%%%%%%%%%%%%%%%%%%%%

Consider the effective quark Lagrangian of strong interactions which is
invariant under a global colour $SU(N_c)$ symmetry
\begin{eqnarray}
\label{enjl}
  {\cal L}&=&\bar{q}(i\gamma^\mu\partial_\mu -\hat{m})q
            +\frac{G_S}{2}[(\bar{q}q)^2+(\bar{q}i\gamma_5\tau_i q)^2]
            \nonumber\\
          &-&\frac{G_V}{2}[(\bar{q}\gamma^\mu\tau_i q)^2
            +(\bar{q}\gamma^\mu\gamma_5\tau_i q)^2].
\end{eqnarray}
Here $q$ is a flavour doublet of Dirac spinors for quark fields $\bar q=(\bar 
u, \bar d)$. Summation over the colour indices is implicit. We use the standard
notation for the isospin Pauli matrices $\tau_i$. The current quark mass matrix
$\hat{m}=\mbox{diag}(m_u, m_d)$ is chosen in such a way that $m_u=m_d$. Without
this term the Lagrangian (\ref{enjl}) would be invariant under global chiral 
$SU(2)\times SU(2)$ symmetry. The coupling constants $G_S$ and $G_V$ have
dimensions $(\mbox{Length})^2$ and can be fixed from the meson mass spectrum.

The transformation law for the quark fields is the following
\begin{equation}
\label{quark}
   \delta q=i(\alpha +\gamma_5\beta )q, \quad
   \delta\bar{q}=-i\bar{q}(\alpha -\gamma_5\beta )
\end{equation}
where parameters of global infinitesimal chiral transformations are chosen as
$\alpha =\alpha_i\tau_i, \ \ \beta =\beta_i\tau_i$. Therefore our basic
Lagrangian ${\cal L}$ transforms according to the law
\begin{equation}
\label{sb}
   \delta {\cal L}=-2i\hat{m}(\bar{q}\gamma_5\beta q).
\end{equation}
It is clear that nothing must destroy this symmetry breaking requirement of the
model (we are not considering anomalies here).

Following the standard procedure we introduce colour singlet collective
bosonic fields in such a way that the action becomes bilinear in the quark
fields and the quark integration becomes trivial
\begin{eqnarray}
\label{gf1}
   Z&=&\int {\cal D}q{\cal D}\bar{q}{\cal D}s{\cal D}p_i{\cal D}V_\mu^i
      {\cal D}A_\mu^i
      \mbox{exp}\left\{i\int d^4x\left[{\cal L} \right.\right.\nonumber \\
    &-&\left.\left.\frac{1}{2G_S}(s^2+p_i^2)+\frac{1}{2G_V}
      (V_{\mu i}^2+A_{\mu i}^2)\right]\right\}.
\end{eqnarray}
We suppress external sources in the generating functional $Z$ and assume 
summation over repeated Lorentz $(\mu )$ and isospin $(i=1,2,3)$ indices. One 
has to require from the new collective variables that
\begin{equation}
\label{req}
   \delta (s^2+p_i^2)=0, \quad
   \delta (V_{\mu i}^2+A_{\mu i}^2)=0
\end{equation}
in order not to destroy the symmetry of the basic Lagrangian ${\cal L}$.
After replacement of variables in $Z$
\begin{equation}
   s=\sigma -\hat{m}+G_S(\bar{q}q),
\end{equation}
\begin{equation}
   p_i=\pi_i -G_S(\bar{q}i\gamma_5\tau_iq),
\end{equation}
\begin{equation}
   V_{\mu}^i=v_\mu^i+G_V(\bar{q}\gamma_\mu\tau_iq),
\end{equation}
\begin{equation}
   A_{\mu}^i=a_\mu^i+G_V(\bar{q}\gamma_\mu\gamma_5\tau_iq),
\end{equation}
these requirements together with (\ref{quark}) lead to the transformation laws
for the new collective fields
\begin{equation}
\label{st}
   \delta\sigma =-\{\beta, \pi\}, \quad
   \delta\pi =i[\alpha, \pi ]+2(\sigma -\hat{m})\beta ,
\end{equation}
\begin{equation}
\label{transva}
   \delta v_\mu =i[\alpha, v_\mu ]+i[\beta , a_\mu ], \quad
   \delta a_\mu =i[\alpha, a_\mu ]+i[\beta , v_\mu ].
\end{equation}
We have introduced the notation $\pi=\pi_i\tau_i$, $v_\mu=v_{\mu i}\tau_i$, 
$a_\mu=a_{\mu i}\tau_i$. Therefore the transformation law of the quark fields 
finally defines the transformation law of the bosonic fields.

The Lagrangian in the new variables (${\cal L}\rightarrow {\cal L}'$) has the 
form
\begin{equation}
\label{lq}
  {\cal L}'=\bar{q}Dq
           -\frac{(\sigma -\hat{m})^2+\pi_i^2}{2G_S}
           +\frac{v_{\mu i}^2+a_{\mu i}^2}{2G_V}.
\end{equation} 
where 
\begin{equation}
\label{D}
              D=i\gamma^\mu\partial_\mu -\sigma +i\gamma_5\pi 
               +\gamma^\mu (v_\mu +\gamma_5 a_\mu ). 
\end{equation}
Let us note that although the Dirac operator $D$ does not include the current
quark mass, $\hat{m}$, the transformation law of pion fields does. Thus,
\begin{equation}
   \delta D=i[\alpha ,D]-i\{\gamma_5\beta ,D\}-2i\hat{m}\gamma_5\beta ;
\end{equation}
i.e., the transformation law of the Dirac operator has an inhomogeneous term
which is proportional to $\hat{m}$. In particular, we have
\begin{equation}
\label{dddd}
   \delta (D^{\dagger}D)=i[\alpha +\gamma_5\beta ,D^\dagger D]
                        +2i\hat{m}(\gamma_5\beta D-D^{\dagger}\gamma_5\beta ).
\end{equation}
The second term $\sim\hat{m}$ can be used to get systematically the explicit 
symmetry breaking pattern of the effective Lagrangian derived on the basis of 
formula (\ref{logdet}). The simplest way to do this is to work in Euclidean 
space. Since the Dirac $\gamma$-matrices in this space are antihermitian the 
combination proportional to the derivatives contained in $(\gamma_5\beta 
D-D^{\dagger}\gamma_5\beta)$ will vanish. It simplifies substantially the 
evaluation of
$\delta\ln |\det D|$ and allows to derive in a closed form the
functional $P$ in (\ref{newlndet}). After that the asymptotic expansion of $P$
to obtain the correcting polynomials at each power of the proper-time
will be a purely technical procedure. However in this paper we prefer to work 
directly in 
Minkowski space and present an alternative way to derive correcting polynomials
step by step starting from the first term of the proper-time expansion.

The subsequent integration over quark fields shows that the effective potential
has a non-trivial minimum and that spontaneous chiral symmetry breaking takes
place. Redefining the scalar field $\sigma\rightarrow\sigma +m$ we come finally
to the effective action
\begin{equation}
\label{seff}
   S_{\mbox{eff}}=-i\ln\det D_m-\int d^4x
   \left[\frac{(\sigma +m-\hat{m})^2+\pi_i^2}{2G_S}-
         \frac{v_{\mu i}^2+a_{\mu i}^2}{2G_V}\right]
\end{equation}
where the Dirac operator $D_m$ is equal to
\begin{equation}
   D_m=i\gamma^\mu\partial_\mu -m-\sigma +i\gamma_5\pi +
       \gamma^\mu (v_\mu +\gamma_5 a_\mu ).
\end{equation}
In this broken phase the transformation law of the pion field changes to 
\begin{equation}
\label{deltapi}
   \delta\pi =i[\alpha, \pi ]+2(\sigma +m-\hat{m})\beta
\end{equation}
in full agreement with the variable replacement $\sigma\rightarrow\sigma +m$ 
for the scalar field in (\ref{st}). What remains to be done to have an explicit
representation of the effective action to leading order in the low energy 
expansion is to evaluate the determinant of the differential operator $D_m$.
We shall consider this problem in the following section.

%%%%%%%%%%%%%%%%%%%%%%%%%%%%%%%%%%%%%%%%%%%%%%%%%%%%%%%%%%%%%%%%%%%%%%%%%%%%%%%
\section{Calculation of the real part of the fermion determinant: the current
         quark mass effect}
%%%%%%%%%%%%%%%%%%%%%%%%%%%%%%%%%%%%%%%%%%%%%%%%%%%%%%%%%%%%%%%%%%%%%%%%%%%%%%%

The modulus of the fermion determinant, $\ln |\det D_m|$, is conveniently 
calculated using the heat kernel method or even more directly in the way 
suggested in \cite{Ball:1989}. The result of these calculations on the basis of
formula (\ref{logdet}) is well known, see for example \cite{Ebert:1986}. We 
prefer this way to the direct calculation of Feynman one-loop integrals 
\cite{Eguchi:1976,Ebert:1982,Volkov:1984,Klimt:1990}, since we need a method 
which allows to control the symmetry content of the result at each considered 
step. The differential operator $D_m$ depends on collective meson fields which 
have well defined transformation laws with respect to the action of the chiral 
group. If one neglects the current quark mass term in the basic quark 
Lagrangian the combination $D_m^\dagger D_m$ transforms covariantly, i.e. 
\begin{equation}
\label{dcov}
   \delta (D_m^{\dagger}D_m)=i[\alpha +\gamma_5\beta ,D_m^\dagger D_m].
\end{equation}
This fact ensures that the definition of the real part of $\ln\det D_m$ in 
terms of the proper-time integral (\ref{logdet}) cannot destroy the symmetry 
properties of the basic Lagrangian. However, if $\hat{m}\neq 0$ this is no
longe true. There is no doubt that the current quark mass does break chirality
in the definition $\ln |\det D_m|$, for we have seen in Sec.2 that the 
combination
$D_m^\dagger D_m$ transforms inhomogeneously. The question is however, whether
one should trust the result obtained through the formula
$\ln |\det D_m|$. We have found that this definition needs to be corrected in 
the presence of the explicit symmetry breaking term, since 
otherwise the transformation law of the effective bosonized meson Lagrangian 
will be different from the transformation law of the basic quark Lagrangian, 
i.e. the content of the theory will be changed. As we already mentioned in the
introduction the problem can be solved if we define the real part of the 
fermion determinant through formula (\ref{newlndet}). This definition can be
extended to include the case with the heat kernel suggested by Gasser and 
Leutwyler \cite{Gasser:1984} or vice versa, for the formal part of these
definitions is the same. The question how to extend the Gasser and Leutwyler's
treatment of the chiral fermion determinant to the case of non-renormalizable 
models like NJL has been considered in \cite{Osipov:2000b}. Therefore we 
proceed from the definition
\begin{equation}
\label{lndetbar}
   -i\ln\det D_m=\frac{i}{2}\int^\infty_0\frac{dT}{T}\rho (T,\Lambda^2)
                 \mbox{Tr}\left(e^{-T\bar{D}_m^2}\right)-\int d^4x 
                 P(\sigma ,\pi ,v_\mu ,a_\mu )
\end{equation}
where $P$ picks up all inessential contributions contained in the proper time 
integral including the terms with the explicit symmetry breaking. The operator 
$\bar{D}_m$ is of the form $\bar{D}_m=\gamma_5 D_m$. At this level
one should not worry that the expression $\bar{D}_m^2$ does not transform 
covariantly under the action of the chiral group. There is nothing wrong with 
this, as long as one is careful to express the final result in terms of chiral 
invariant quantities. The present procedure allows to do it in a systematic and
consistent way for each order of the heat kernel expansion. The functional 
$P(\sigma ,\pi ,v_\mu ,a_\mu )$ depends on collective fields and their 
derivatives. We define 
it by requiring the real part of the fermion determinant to transform as 
Lagrangian (\ref{enjl}). The imaginary part of $\ln\det D_m$ will be discussed 
elsewhere. The expression (\ref{lndetbar}) belongs to the ones which are known 
as proper-time regularizations. In the case of non-renormalizable models like 
ENJL we have to introduce the cutoff $\Lambda$ to render the integrals over 
$T$ convergent. We consider a class of regularization schemes which can be 
incorporated in the expression (\ref{lndetbar}) through the kernel 
$\rho (T,\Lambda^2)$. These regularizations allow to shift in loop momenta. A 
typical example is the covariant Pauli-Villars cutoff \cite{Pauli:1949}
\begin{equation}
\label{cc}
      \rho (T, \Lambda^2)=1-(1+T\Lambda^2)e^{-T\Lambda^2}.
\end{equation}

Let us put this expression into formula (\ref{logdet}) and calculate the 
corresponding effective potential $V(\sigma ,\pi_i )$, using eq.(\ref{D}) with
fields $v_\mu$ and $a_\mu$ switched off. We have as a result that
\begin{eqnarray}
\label{effpot} 
\lefteqn{
   V(\sigma ,\pi_i )=\frac{\hat{m}\sigma}{G_S}+\frac{\sigma^2+\pi^2_i}{2G_S}
            \left(1-\frac{N_cG_S\Lambda^2}{4\pi^2}\right)
            } \nonumber\\
            &+&\frac{N_c}{8\pi^2}\left[(\sigma^2+\pi^2_i)^2\ln\left(1+
               \frac{\Lambda^2}{\sigma^2+\pi^2_i}\right)-\Lambda^4\ln\left(1+
               \frac{\sigma^2+\pi^2_i}{\Lambda^2}\right)\right].
\end{eqnarray}
The minimum of this potential is localized at the point $\sigma =<\sigma >_0=m$
which is the solution of the ``gap"-equation
\begin{equation}
\label{gap}
  \frac{m-\hat{m}}{mG_S}=\frac{N_cJ_0}{2\pi^2}.
\end{equation}
The function $J_0$ is one of the set of integrals $J_n$ appearing in the 
result of the asymptotic expansion of (\ref{lndetbar})
\begin{equation}
J_n=\int^\infty_0\frac{dT}{T^{2-n}}e^{-Tm^2}\rho (T,\Lambda^2),
         \quad n=0,1,2...
\end{equation}
Although the potential $V(\sigma ,\pi_i )$ leads to the correct form of the 
``gap"-equation\footnote{It is the same solution as the one from the 
Schwinger-Dyson equation.} \cite{Nambu:1961}, it is incomplete in its 
$\hat{m}$-dependent part. The reason is obvious: it destroys the symmetry 
breaking pattern of the basic Lagrangian, as one can conclude after short 
calculations. We did not include in (\ref{effpot}) the corresponding part 
from the functional $P$.  

Let us show how to get these counterterms on the basis of formula 
(\ref{lndetbar}). We have for $\bar{D}_m^2$ the following representation
\begin{equation}
\label{D2}
   \bar{D}_m^2=d^\mu d_\mu +m^2+Q
\end{equation}
where
\begin{eqnarray}
\label{Q} 
  d_\mu&=&\partial_\mu +A_\mu ,\quad 
          A_\mu =\gamma^\mu\gamma_5\pi -iv_\mu 
          +\frac{i}{2}[\gamma^\nu ,\gamma^\mu ]\gamma_5 a_\nu ,\nonumber \\
  Q&=&\sigma^2+2m\sigma +3\pi^2-2a^\mu a_\mu +i\gamma^\mu
      \left(\partial_\mu\sigma +2\{a_\mu ,\pi\}\right)\nonumber \\
   &-&\frac{1}{2}[\gamma^\mu ,\gamma^\nu ](a_\mu a_\nu +v_\mu v_\nu 
      +i\partial_\mu v_\nu )\nonumber \\
   &-&i\gamma_5\left(\partial_\mu a^\mu +i[a_\mu ,v^\mu ]
      +2(\sigma +m)\pi\right).
\end{eqnarray}

The functional trace in formulae (\ref{lndetbar}) is equal to
\begin{equation}
\label{ft}
  \mbox{Tr}\left(e^{-T\bar{D}_m^2}\right)
  =i\int d^4x\frac{e^{-Tm^2}}{(4\pi T)^2}
  \sum_{n=0}^{\infty}\mbox{tr}(T^na_n)
\end{equation}
where tr denote the traces over colour, flavour and Lorentz indices.
The coefficients $a_n\equiv a_n(x,x)$ are the coincidence limit of Seeley --
DeWitt coefficients. We need the first three of them for our purposes
\begin{equation}
\label{sdw}
  a_0=1,\quad a_1=-Q,\quad a_2=\frac{1}{2}Q^2+\frac{1}{12}F^2
\end{equation}
where $F^2=F^{\mu\nu}F_{\mu\nu}$ and $F_{\mu\nu}=[d_\mu ,d_\nu ]$.

In \cite{Osipov:2000b} we have shown for $\hat{m}=0$ how to obtain the Gasser 
and Leutwyler's part of functional $P$ by which this definition of heat kernel
needs to be modified in order to arrive at the fermion determinant whose real 
part is invariant under chiral transformations. Let us now show how to get the 
explicit symmetry breaking part $P'$ of the functional $P$. Restricting to the 
second order Seeley -- DeWitt coefficient one can obtain from (\ref{seff}) and 
(\ref{lndetbar}) the effective Lagrangian
\begin{eqnarray}
\label{leff}
   {\cal L}_{\mbox{eff}}
   &=&\frac{v_{\mu i}^2+a_{\mu i}^2}{2G_V}-\frac{1}{2G_S}
     [(\sigma +m-\hat{m})^2+\pi^2_i ]+\frac{N_cJ_0}{4\pi^2}(\sigma^2+2m\sigma
     +\pi^2_i) \nonumber \\
   &-&\frac{N_cJ_1}{8\pi^2}\left[\frac{1}{6}\mbox{tr}(v_{\mu\nu}^2
     +a_{\mu\nu}^2)-\frac{1}{2}\mbox{tr}\left((\nabla'_\mu\pi )^2
     +(\nabla_\mu\sigma )^2\right)\right.\nonumber \\
   &+&\left.(\sigma^2+2m\sigma +\pi^2_i)^2\right]
     -P'(\sigma ,\pi ,v_\mu ,a_\mu )
\end{eqnarray}
where trace is to be taken in isospin space. In the considered approximation
the functional $P'$ is simply a polynomial. Here we have used the notation
\begin{equation}
   v_{\mu\nu}=\partial_\mu v_\nu -\partial_\nu v_\mu -i[v_\mu ,v_\nu ]
             -i[a_\mu ,a_\nu ],
\end{equation}     
\begin{equation}
   a_{\mu\nu}=\partial_\mu a_\nu -\partial_\nu a_\mu -i[a_\mu ,v_\nu ]
             -i[v_\mu ,a_\nu ],
\end{equation}     
\begin{equation}
   \nabla_\mu\sigma =\partial_\mu\sigma -i[v_\mu ,\sigma ]+\{a_\mu ,\pi\},
\end{equation}    
\begin{equation}
   \nabla'_\mu\pi =\partial_\mu\pi -i[v_\mu ,\pi ]-\{a_\mu ,\sigma +m\}.
\end{equation}

In formula (\ref{leff}) we already fixed the part of the functional $P$ which 
is responsible for the chiral symmetric contribution. Now we need only to
determine the explicit symmetry breaking part $P'(\sigma ,\pi ,v_\mu ,a_\mu )$.
If one uses the classical equation of motion for the pion field, 
$\pi_i=iG_S\bar{q}\gamma_5\tau_i q$, see (\ref{lq}), one can rewrite
eq.(\ref{sb}) in terms of meson fields
\begin{equation}
\label{sbm}
   \delta {\cal L}=-\frac{2\hat{m}}{G_S}(\beta_i\pi_i ).
\end{equation}

Now our task is to choose the polynomial $P'$ in such a way that the 
Lagrangian (\ref{leff}) will have the same transformation law. Let us note that
$P'$ is unique up to a chirally invariant polynomial. One can always choose 
$P'$ in such a manner that the ``gap"-equation is not modified, i.e. using this
chiral symmetry freedom to avoid from $P'$ terms linear in $\sigma$. One can do
this noting that $\delta (\sigma^2+\vec{\pi}^2)=-2(m-\hat{m})\delta\sigma$. It
completely fixes the chiral freedom in $P'$. The
variation of $P'(\sigma ,\pi ,v_\mu ,a_\mu )$ has to cancel all terms which 
break explicitly chiral symmetry, excluding (\ref{sbm}). As a result we get that
\begin{eqnarray}
\lefteqn{
  P'(\sigma ,\pi ,v_\mu ,a_\mu )
  =\frac{\hat{m}^2(\sigma^2+\pi_i^2 )}{2m(m-\hat{m})G_S}
    -\hat{m}\frac{N_cJ_1}{2\pi^2}[(2m-\hat{m})\sigma^2+\sigma
    (\sigma^2+\pi_i^2 )]
        } \nonumber \\
  & &+\hat{m}\frac{N_cJ_1}{4\pi^2}\mbox{tr}\{(2m-\hat{m})a^2_\mu
     -a_\mu\partial_\mu\pi +ia_\mu [v_\mu ,\pi ]+2\sigma a^2_\mu\}.
\end{eqnarray}

Finally we have the following expression for the effective Lagrangian
\begin{eqnarray}
\label{fleff}
   {\cal L}_{\mbox{eff}}
   &=&\frac{v_{\mu i}^2+a_{\mu i}^2}{2G_V}-\frac{\hat{m}(\sigma^2 
     +\vec{\pi}^2)}{2(m-\hat{m})G_S}
     -\frac{N_cJ_1}{8\pi^2}\left[\frac{1}{6}\mbox{tr}(v_{\mu\nu}^2
     +a_{\mu\nu}^2)\right.\nonumber \\
   &-&\left.\frac{1}{2}\mbox{tr}\left((\nabla_\mu\pi )^2
     +(\nabla_\mu\sigma )^2\right)
     +\left(\sigma^2+2(m-\hat{m})\sigma +\pi^2_i\right)^2\right]
\end{eqnarray}
where
\begin{equation}
   \nabla_\mu\pi =\partial_\mu\pi -i[v_\mu ,\pi ]-\{a_\mu ,\sigma +m-\hat{m}\}.
\end{equation}
It can be verified, by explicit calculation, that the second term in this
expression gives the correct behaviour of the Lagrangian with respect to chiral
transformations. Other terms are combined in chiral invariant groups. For 
example, the terms proportional to $J_1$ are chiral invariant. The same will be
true for each group of terms at the same $J_n$ where $n\geq 1$. In particular
one can in this way obtain systematically and step by step in the expansion 
(\ref{ft}) the $\hat{m}$-part of the effective potential which escaped from 
(\ref{effpot}).

%%%%%%%%%%%%%%%%%%%%%%%%%%%%%%%%%%%%%%%%%%%%%%%%%%%%%%%%%%%%%%%%%%%%%%%%%%%%%%%
\section{$\pi a_\mu$ mixing, field renormalizations and meson mass spectrum}
%%%%%%%%%%%%%%%%%%%%%%%%%%%%%%%%%%%%%%%%%%%%%%%%%%%%%%%%%%%%%%%%%%%%%%%%%%%%%%%

Having established the effective Lagrangian one may look for kinetic and mass 
terms of the composite meson fields and extract the physical meson masses by 
bringing the kinetic terms to the canonical form by means of field 
renormalizations. It is however known that for the axial-vector field $a_\mu$
one encounters the complication that chiral symmetry allows for terms which
induce mixing between $a_\mu$ and pseudoscalar mesons. Such couplings must and
can always be transformed away by a transformation which removes the spin-0
component of $a_\mu$. The simplest replacement of variables which really 
fulfills the necessary transformation property (\ref{transva}) is
\begin{eqnarray}
\label{ared}
  a_\mu&=&a'_\mu +\frac{\kappa}{2}\left(\{\sigma +m-\hat{m}, \partial_\mu\pi\}
         -\{\pi ,\partial_\mu\sigma\}\right),\nonumber \\
  v_\mu&=&v'_\mu +\frac{i\kappa}{2}\left([\sigma , \partial_\mu\sigma ]
         +[\pi ,\partial_\mu\pi ]\right),
\end{eqnarray}
reminiscent of the axial-vector and vector currents of the linear sigma model,
\cite{Itz:1980}. One can see that these formulae include the linear part of 
formula (\ref{wrong}) and do not lead to unwanted linear contributions in the 
vector field transformations (there is no any $\sigma - v_\mu$ mixing in this
model). One can conclude that eq.(\ref{wrong}) is just a piece of a more 
complicated expression which has to be used for a correct removing of the $\pi 
a_\mu$ mixing effect in this approach. In the case under consideration the 
commutator $[\sigma , \partial_\mu\sigma ]=0$. These new redefinitions, as 
compared to (\ref{wrong}), will induce changes at the level of couplings with 
three or more fields. In the Appendix it is shown that the replacement 
(\ref{ared}) is identical to the field redefinition considered in 
\cite{Bijnens:1993} for the case of nonlinear realization of chiral symmetry. 
The constant $\kappa$ is fixed by the requirement that the bilinear part of 
the effective Lagrangian becomes diagonal in the fields $\pi , a'_\mu$. We find
in this way that
\begin{equation}
\label{kappa}
   \frac{1}{2\kappa}=\left(m-\hat{m}\right)^2+\frac{\pi^2}{N_cJ_1G_V}.
\end{equation}

To define the physical meson fields let us consider the bilinear part of the
effective Lagrangian
\begin{eqnarray}
\label{bilpart}
   {\cal L}_{\mbox{free}}
   &=&\frac{1}{4}\mbox{tr}\left\{G_V^{-1}\left[(v'_\mu )^2+(a'_\mu )^2
      +\kappa^2(m-\hat{m})^2(\partial_\mu\pi )^2
      -\frac{\hat{m}(\sigma^2+\pi^2)}{m-\hat{m}}\right]\right. \nonumber \\
   &+&\frac{N_cJ_1}{4\pi^2}\left[(\partial_\mu\sigma )^2
      +g_A^2(\partial_\mu\pi )^2
      +4(m-\hat{m})^2[(a'_\mu)^2-\sigma^2]\right.\nonumber \\
   &-&\left.\left. \frac{1}{3}(\partial_\mu v'_\nu -\partial_\nu v'_\mu )^2
      -\frac{1}{3}(\partial_\mu a'_\nu -\partial_\nu a'_\mu )^2
      \right]\right\}.
\end{eqnarray}
The following renormalizations lead to the standard form for the kinetic terms
of spin-1 fields 

\begin{equation}
\label{renorm1}
   v'_\mu =\sqrt{\frac{6\pi^2}{N_cJ_1}}v_\mu^{(ph)}\equiv
           \frac{g_{\rho}}{2}v_\mu^{(ph)}, 
   \quad   a'_\mu =\frac{g_{\rho}}{2}a_\mu^{(ph)}.
\end{equation}
Then we have
\begin{equation}
\label{mass1}
   m^2_\rho =\frac{6\pi^2}{N_cJ_1G_V}, \quad m^2_a=m^2_\rho +6(m-\hat{m})^2.
\end{equation}
In particular it implies the relations 
\begin{equation}
\label{Z}
   g_A=1-\frac{6(m-\hat{m})^2}{m^2_a}=\frac{m^2_\rho}{m^2_a},
   \quad \kappa =\frac{3}{m^2_a}.
\end{equation}
We also have to redefine the spin-0 fields
\begin{equation}
\label{renorm0}
  \sigma =\sqrt{\frac{4\pi^2}{N_cJ_1}}\sigma^{(ph)}\equiv
          g_{\sigma}\sigma^{(ph)}, \quad \pi =g_\pi\pi^{(ph)}, 
          \quad g_\pi =\frac{g_\sigma}{\sqrt{g_A}}.
\end{equation}
The mass formulae for spin-0 fields are
\begin{equation}
\label{mass0}
   m^2_\pi =\frac{\hat{m}g^2_\pi}{(m-\hat{m})G_S}, \quad 
   m^2_\sigma =m^2_\pi +4(m-\hat{m})^2.
\end{equation}
As compared with previous calculations in 
\cite{Volkov:1984,Ebert:1986,Bijnens:1993} our 
mass formulae have a different dependence on the current quark mass.
Numerically these lead to small deviations in the final results for 
$\hat{m}\sim 7\ \mbox{MeV}$. However in the case of broken $SU(3)\times SU(3)$
symmetry this effect is more essential and has to be taken into account with
all care. Anyway, the small numerical difference between final results cannot 
justify the incorrect treatment of symmetry principles. 

Let us also point out that after the field redefinitions the symmetry breaking 
pattern takes the form \cite{Gas:1969}
\begin{equation}
\label{sbf}
   \delta {\cal L}_{\mbox{eff}}=-2m^2_\pi f_\pi \beta_i\pi_i^{(ph)}
\end{equation}
where we used the relation
\begin{equation}
   g_\pi =\frac{m-\hat{m}}{f_\pi}.
\end{equation}
Formula (\ref{sbf}) leads to the well known PCAC relation for the divergence 
of the quark axial-vector current 
\begin{equation}
\label{pcac}
   \partial_\mu\vec{J}^\mu_5=2f_\pi m^2_\pi \vec{\pi}^{(ph)}.
\end{equation}

%%%%%%%%%%%%%%%%%%%%%%%%%%%%%%%%%%%%%%%%%%%%%%%%%%%%%%%%%%%%%%%%%%%%%%%%%%%%%%%
\section{The coset-space parametrization for meson fields}
%%%%%%%%%%%%%%%%%%%%%%%%%%%%%%%%%%%%%%%%%%%%%%%%%%%%%%%%%%%%%%%%%%%%%%%%%%%%%%%

To compare our Lagrangian in full detail with the result of the nonlinear 
approach \cite{Bijnens:1993} one has to perform a chiral field dependent
rotation which eliminates the nonderivative coupling of $\pi$.
A systematic treatment of the problem has been developed by Coleman, Wess and
Zumino \cite{Coleman:1969}. We consider here the approximation to this picture
which is known as the nonlinear realization in which no scalar particles exist,
i.e. we shall eliminate completely the scalar degree of freedom from the meson 
Lagrangian (\ref{fleff}). The conventional method to realize this idea is based
on the fact that the sum $(\sigma +m-\hat{m})^2+\vec{\pi}^2$ is invariant under
chiral transformations. Therefore one can put it equal to a constant (the 
nonlinear ansatz) without spoiling chiral symmetry. In this case the scalar
field $\sigma$ is no more an independent variable and can be excluded from the
Lagrangian \cite{Gursey:1961,Gas:1969} in favour of the pion field. A constant 
can be fixed at a point $<\sigma >=<\pi_i>=0$, i.e. 
\begin{equation}
\label{constr}
   (\sigma +m-\hat{m})^2+\vec{\pi}^2=(m-\hat{m})^2.  
\end{equation}
The theory with the constraint (\ref{constr}) can be formulated in terms of
the generating functional 
\begin{eqnarray}
\label{gf2}
   Z_1&=&\int {\cal D}q{\cal D}\bar{q}{\cal D}\sigma {\cal D}\pi_i
         {\cal D}v_\mu^i{\cal D}a_\mu^i{\cal D}\lambda
         \mbox{exp}\left\{i\int d^4x\left[{\cal L}_{\mbox{qm}} 
         \right.\right.\nonumber \\
      &-&\left.\left.\frac{\lambda}{2}\left(
         (\sigma +m-\hat{m})^2+\vec{\pi}^2-(m-\hat{m})^2\right)  
         \right]\right\}
\end{eqnarray}
where ${\cal L}_{\mbox{qm}}$ is the quark-meson Lagrangian (\ref{lq}) rewritten
in the broken phase (the replacement $\sigma\rightarrow\sigma +m$ is done).
One has to take an integral over the Lagrange multiplier $\lambda (x)$. It
leads to the $\delta$-functional
\begin{equation}
   \delta\left[(\sigma +m-\hat{m})^2+\vec{\pi}^2-(m-\hat{m})^2\right]=
   \sum^2_{a=1}\frac{\delta (\sigma -\sigma_a)}{
   2\sqrt{(m-\hat{m})^2-\vec{\pi}^2}           }
\end{equation}
where
\begin{equation}
  \sigma_{1,2}=(m-\hat{m})\left(\pm\sqrt{1-\frac{\vec{\pi}^2}{(m-\hat{m})^2}}
              -1\right).
\end{equation}
In $Z_1$ one has to integrate only the configurations $\sigma$ with $<\sigma 
>=0$. It means that only $\sigma_1$ contributes to the generating functional, 
for $<\sigma_2 >\neq 0$. After integrating over $\sigma$ and quark fields we 
have 
\begin{equation}
\label{gf3}
   Z_1=\int {\cal D}\mu [\pi_i] {\cal D}v_\mu^i{\cal D}a_\mu^i 
       \mbox{exp}\left\{i\int d^4x {\cal L}'_{\mbox{eff}}(
       \pi_i,v_\mu^i,a_\mu^i)                                            
       \right\}.
\end{equation}
The Lagrangian ${\cal L}'_{\mbox{eff}}(\pi_i,v_\mu^i,a_\mu^i)$ is our 
Lagrangian (\ref{fleff}) where one has to substitute $\sigma$ by $\sigma_1$.
The $SU(2)\times SU(2)$-invariant measure ${\cal D}\mu [\pi_i]$ emerging in
$Z_1$ is related with the curvature in the space of $\pi_i$ variables,  
\begin{equation}
   \prod^3_{i=1}{\cal D}\mu [\pi_i]=\prod^3_{i=1}\frac{{\cal D}\pi_i}{
                        \sqrt{1-\frac{\vec{\pi}^2}{(m-\hat{m})^2}}}.
\end{equation}
It is more convenient now to introduce new variables $\phi_i$, different from 
$\pi_i$, defined as 
\begin{equation}
\label{phi}
   \pi_i=(m-\hat{m})\frac{\phi_i}{\phi}\sin\phi , \quad \phi =\sqrt{\phi_i^2}.
\end{equation}
The parametrization in terms of $\phi_i$ fields corresponds to the normal 
coordinate system on the surface of a three-dimensional sphere: the $SU(2)
\times SU(2)/SU(2)$ group manifold. One can get the expression for the
invariant measure in these new variables
\begin{equation}
   \prod^3_{i=1}{\cal D}\mu [\pi_i]=\prod^3_{i=1}\det\left(
                 \frac{\partial\pi_n}{\partial\phi_m}\right)
                 \frac{{\cal D}\phi_i}{\cos\phi}
       =N\prod^3_{i=1}\frac{\sin^2\phi}{\phi^2}{\cal D}\phi_i .      
\end{equation}
We drop here the expression for the nonessential factor $N$. 

It is not difficult to get from (\ref{deltapi}) the form of the infinitesimal 
chiral transformation for the variables $\phi_i$
\begin{equation}
   \delta\phi_i=2\beta_i\phi\cot\phi +2\frac{\phi_i\phi_k}{\phi^2}\beta_k
                (1-\phi\cot\phi )-2\epsilon_{ijk}\alpha_j\phi_k .  
\end{equation}
Geometrically this transformation is nothing else than a law for the coordinate
changes in the $SU(2)\times SU(2)/SU(2)$ coset-space under the action of
the $SU(2)\times SU(2)$ chiral group. Let us rewrite the Lagrangian 
${\cal L}'_{\mbox{eff}}(\pi_i,v_\mu^i,a_\mu^i)$ in terms of
these new variables. It is customary to put it in the form which includes the
fields with the covariant transformation law. For this purpose one has to
introduce also the new vector and axial-vector variables
\begin{eqnarray}
\label{newf1}
   v_\mu&=&\frac{1}{2}\left[\xi^\dagger (v'_\mu +a'_\mu )\xi
                            +\xi (v'_\mu -a'_\mu )\xi^\dagger\right] 
           \nonumber \\
   a_\mu&=&\frac{1}{2}\left[\xi^\dagger (v'_\mu +a'_\mu )\xi
                            -\xi (v'_\mu -a'_\mu )\xi^\dagger\right]. 
\end{eqnarray}
We use here the standard definition of the coset representative $\xi$
\begin{equation}
\label{xi}
   \xi =\mbox{exp}\left(-\frac{i}{2}\tau_i\phi_i\right).
\end{equation}                                           
One can show that the Jacobian of this replacement is equal to one
\begin{equation}
   \frac{\partial (v_{i\mu},a_{i\mu})}{\partial (v'_{j\mu},a'_{j\mu})}=1
\end{equation}
and the transformation laws of new $v'_\mu ,a'_\mu$ fields are covariant, i.e.
\begin{equation}
   \delta v'_\mu =i[\alpha +\beta (\phi ), v'_\mu ], \quad
   \delta a'_\mu =i[\alpha +\beta (\phi ), a'_\mu ] 
\end{equation}
where
\begin{equation}
   \beta (\phi )=\beta_k(\phi )\tau_k, \quad 
   \beta_k(\phi )=\epsilon_{kni}\beta_i\frac{\phi_n}{\phi}\tan\frac{\phi}{2}.
\end{equation}
Let us note that the part of chiral transformations which depends on the 
parameter $\beta$ is $x$-dependent now, $\beta (\phi )$, through the field 
$\phi (x)$. We remind also with the purpose of future references that 
the function $\xi_\mu$ has the same transformation law  
\begin{equation}
   \xi_\mu =-i(\xi\partial_\mu\xi^\dagger -\xi^\dagger\partial_\mu\xi ), 
   \quad  \delta\xi_\mu =i[\alpha +\beta (\phi ), \xi_\mu ]. 
\end{equation}                                               
Another function, $\Gamma_\mu$, defines the covariant derivative in the coset 
space, i.e. if $R$ transforms covariantly than the same is true for its
covariant derivative $d_\mu R$ defined by
\begin{equation}
   d_\mu R=\partial_\mu R+[\Gamma_\mu , R]
\end{equation}
where
\begin{equation}
   \Gamma_\mu =\frac{1}{2}\left(\xi\partial_\mu\xi^\dagger 
              +\xi^\dagger\partial_\mu\xi\right). 
\end{equation} 
One can show that $\Gamma_\mu$ transforms like Yang -- Mills connection on the 
given coset space
\begin{equation}
   \delta\Gamma_\mu =i[\alpha +\beta (\phi ), \Gamma_\mu ]
                    -i\partial_\mu\beta (\phi ). 
\end{equation}                                               
All these functions appear naturally in the effective Lagrangian (\ref{gf3})
by means of the abovementioned replacements of variables and we have as a
result
\begin{eqnarray}
\label{nonl}
   {\cal L}'_{\mbox{eff}}&=&\frac{1}{4G_V}\mbox{tr}\left[(v'_\mu )^2
            +(a'_\mu )^2\right]+\frac{m^2_\pi}{4}f_\pi^2\mbox{tr}\left(\xi\xi 
            +\xi^\dagger\xi^\dagger -2\right) \nonumber \\
   &+&\frac{N_cJ_1}{16\pi^2}\left[(m-\hat{m})^2\mbox{tr}(\xi_\mu -2a'_\mu )^2
      -\frac{1}{3}\mbox{tr}(\tilde{V}_{\mu\nu}^2+\tilde{A}_{\mu\nu}^2)\right]
\end{eqnarray}
where we put
\begin{equation}
   \tilde{V}_{\mu\nu}={\cal V}'_{\mu\nu}+\frac{i}{2}\left([\xi_\mu ,a'_\nu ]
                                          -[\xi_\nu ,a'_\mu ]\right),
\end{equation}
\begin{equation}
   \tilde{A}_{\mu\nu}={\cal A}'_{\mu\nu}+\frac{i}{2}\left([\xi_\mu ,v'_\nu ]
                                          -[\xi_\nu ,v'_\mu ]\right),
\end{equation}
and
\begin{equation}
   {\cal V}'_{\mu\nu}=d_\mu v'_\nu -d_\nu v'_\mu -i[v'_\mu ,v'_\nu ]
                                            -i[a'_\mu ,a'_\nu ],
\end{equation}
\begin{equation}
   {\cal A}'_{\mu\nu}=d_\mu a'_\nu -d_\nu a'_\mu -i[a'_\mu ,v'_\nu ]
                                            -i[v'_\mu ,a'_\nu ].
\end{equation}
 
Similary to the case with linear realization of chiral symmetry one has now
to diagonalize the pseudoscalar -- axial-vector bilinear form in the Lagrangian
(\ref{nonl}). The replacement 
\begin{equation}
\label{repl}
   a'_\mu\rightarrow a'_\mu +\kappa (m-\hat{m})^2\xi_\mu 
\end{equation} 
with $\kappa$ defined by (\ref{kappa}) solves the problem. We have as a result
\begin{eqnarray}
\label{nonl2}
   {\cal L}'_{\mbox{eff}}&=&\frac{f^2_\pi}{4}\mbox{tr}(\xi_\mu\xi_\mu )
            +\frac{m^2_\pi}{4}f_\pi^2\mbox{tr}\left(\xi\xi 
            +\xi^\dagger\xi^\dagger -2\right) \nonumber \\
           &-&\frac{1}{2g^2_\rho}\mbox{tr}(V_{\mu\nu}^2+A_{\mu\nu}^2)
            +\frac{1}{g^2_\rho}\mbox{tr}
              \left[m^2_\rho (v'_\mu )^2+m^2_a (a'_\mu )^2\right] 
\end{eqnarray}
where after the replacement (\ref{repl}) the antisymmetric tensors $V_{\mu\nu}$
and $A_{\mu\nu}$ read
\begin{equation}
   V_{\mu\nu}={\cal V}'_{\mu\nu}
             +\frac{ig_A}{2}\left([\xi_\mu ,a'_\nu ]-[\xi_\nu ,a'_\mu ]\right)
             +\frac{i}{4}(1-g_A^2)[\xi_\mu ,\xi_\nu ],
\end{equation}
\begin{equation}
   A_{\mu\nu}={\cal A}'_{\mu\nu}
             +\frac{ig_A}{2}\left([\xi_\mu ,v'_\nu ]-[\xi_\nu ,v'_\mu ]\right)
             +\frac{1}{2}(1-g_A)(d_\mu\xi_\nu -d_\nu\xi_\mu ).
\end{equation}
The following redefinitions lead us to the physical pseudoscalar, vector, and
axial-vector states
\begin{equation}             
   \phi_i =\frac{1}{f_\pi}\pi_i^{(ph)}, \quad 
   v'_\mu =\frac{g_\rho}{2}v_\mu^{(ph)}, \quad
   a'_\mu =\frac{g_\rho}{2}a_\mu^{(ph)}.
\end{equation} 

The Lagrangian (\ref{nonl2}) is the result of the nonlinear realization in which
no scalar particles exist. It is sufficient to illustrate our point, although
it is an approximation. The first term in (\ref{nonl2}) is the canonical 
Lagrangian for the nonlinear sigma model. The second term of the Lagrangian 
breakes chiral symmetry and obviously satisfies the symmetry breaking pattern 
of the basic quark Lagrangian. Except for that term the rest of the terms in 
(\ref{nonl2}) are manifestly chiral invariant. It means that all 
$\hat{m}$-dependence is absorbed in coupling constants, for instance the
mass of the axial-vector meson, $m_a$, and the coupling, $g_A$, depend on
$\hat{m}$ (see eq.(\ref{mass1}) and eq.(\ref{Z})). Our expression (\ref{nonl2})
cleary shows that only the symmetry breaking part of the Lagrangian includes
the cluster $\Sigma =\xi^\dagger\xi^\dagger +\xi\xi$ with a noncovariant 
transformation law. This combination (and the other similar one: $\Delta =
\xi^\dagger\xi^\dagger -\xi\xi$) never appears among the interaction vertices,
it would generate spurious symmetry breaking effects. The structure of our 
Lagrangian differs in this respect from the known expressions where the 
explicit symmetry breaking effect has also been included (see for instance 
\cite{Espriu:1990,Bijnens:1993}).

%%%%%%%%%%%%%%%%%%%%%%%%%%%%%%%%%%%%%%%%%%%%%%%%%%%%%%%%%%%%%%%%%%%%%%%%%%%%%%%
\section{Concluding remarks}
%%%%%%%%%%%%%%%%%%%%%%%%%%%%%%%%%%%%%%%%%%%%%%%%%%%%%%%%%%%%%%%%%%%%%%%%%%%%%%%

The modulus of the chiral fermion determinant is well defined by the formula
$\ln |\det D|$. It has generally been assumed that this formula can be also 
used in the case when chiral symmetry is explicitly broken. In this work we
have shown in considerable detail that it is not true. We have described a
practical tool to derive in a systematic and consistent way the real part of
$\ln\det D$ considering as an example the ENJL model with the chiral
$SU(2)\times SU(2)$ symmetric four-quark interactions. In this special case we
arrive at the fermion determinant as a result of integration over quark fields
in the corresponding generating functional. The effective meson Lagrangian 
describing the dynamics of collective degrees of freedom emerges in this way. 
The symmetry breaking pattern of the starting quark Lagrangian should not be
changed during bosonization. This symmetry requirement together with the
Schwinger -- Dyson equation which defines the vacuum state of the model helps
us to fix completely the $\hat{m}$-dependent part of the effective Lagrangian.
It differs from the ones obtained on the ground of other methods like the 
direct calculation of the one-loop Feynman diagrams, or the naive use of the
proper-time representation in form (\ref{logdet}). 

In this work we have addressed another point also related with the chiral
symmetry transformation laws. It is associated with the well-known mixing
terms in the pseudoscalar-axialvector fields. We have shown that the way 
presented in the literature to diagonalize this admixture in the case of the 
linear realization of chiral symmetry is not compatible with the transformation
laws for the vector and axial-vector fields and induces spurious symmetry 
breaking terms. We have derived the minimal form of covariant 
redefinitions for the spin-1 fields needed to fulfill simultaneously the 
mesonic transformation laws and diagonalization. 
We find out that our redefinitions of vector and axial-vector fields is an
agreement with the standard redefinition of axial-vector fields in the
nonlinear case.

In the end we have rewritten our Lagrangian in the nonlinear form  
exluding completely the scalar field. This approximation is sufficient to 
pin down the general structures containing the explicit symmetry breaking terms
in the effective mesonic Lagrangian for the ENJL model with  
nonlinear realization of chiral symmetry. We conclude that the effect of
explicit chiral symmetry breaking has never been treated with enough care in
the framework of the NJL model and have presented in this work a consistent 
method to take it into account.

%%%%%%%%%%%%%%%%%%%%%%%%%%%%%%%%%%%%%%%%%%%%%%%%%%%%%%%%%%%%%%%%%%%%%%%%%%%%%%%
\vspace{5.0mm}
\noindent
{\bf Acknowledgements}
\vspace{5.0mm}
%\appendix{Acknowledgements}
%%%%%%%%%%%%%%%%%%%%%%%%%%%%%%%%%%%%%%%%%%%%%%%%%%%%%%%%%%%%%%%%%%%%%%%%%%%%%%%

This work is supported by grants provided by Funda\ca o para a Ci\^encia e a
Tecnologia, PRAXIS/C/FIS/12247/1998, PESO/P/PRO/15127/1999 and NATO
"Outreach" Cooperation Program.

%%%%%%%%%%%%%%%%%%%%%%%%%%%%%%%%%%%%%%%%%%%%%%%%%%%%%%%%%%%%%%%%%%%%%%%%%%%%%%%
\vspace{5.0mm}
\noindent
{\bf Appendix}
\vspace{5.0mm}
%\appendix{Appendix}
%%%%%%%%%%%%%%%%%%%%%%%%%%%%%%%%%%%%%%%%%%%%%%%%%%%%%%%%%%%%%%%%%%%%%%%%%%%%%%%

Let us show here that the fields redefinition (\ref{ared}) coinsides with a
similar redefinition which has been used in the case of non-linear realization
of chiral symmetry. We shall start from the piece of Lagrangian (\ref{lq}) with
the collective fields of spin-1
\begin{eqnarray}
\lefteqn{\bar{q}\gamma^\mu (v_\mu +\gamma_5 a_\mu )q= } \nonumber \\
  & &\bar{q}\gamma^\mu\left\{v'_\mu +i\frac{\kappa}{2}[\pi ,\partial_\mu\pi ]
     +\gamma_5\left[a'_\mu +\kappa\left( (\sigma +m-\hat{m})\partial_\mu\pi 
     -\pi\partial_\mu\sigma\right)\right]\right\}q.
\end{eqnarray}
To come to the non-linear realization of chiral symmetry one has to eliminate 
the scalar field, which is achieved by the constraint
\begin{equation}
   \sigma +m-\hat{m}=(m-\hat{m})\sqrt{1-\frac{\pi^2_i}{(m-\hat{m})^2}}.
\end{equation}
Let us choose the exponentional parametrization for pion fields
\begin{equation}
   \pi_i=(m-\hat{m})\frac{\phi_i}{\phi}\sin\phi , \quad \phi =\sqrt{\phi_i^2}.
\end{equation}
The pure geometrical picture appears if we redefine at the same time the quark
fields
\begin{equation}
   Q=(\xi^\dagger P_R+\xi P_L)q, \quad \xi =\mbox{exp}
     \left(-\frac{i}{2}\tau_i\phi_i\right)
\end{equation}                                           
where the projection operators $2P_R=(1+\gamma_5)$ and $2P_L=(1-\gamma_5)$ have
been introduced. In this case one needs also to redefine the vector and 
axial-vector fields. The new variables $V_\mu$ and $A_\mu$ are the following
ones
\begin{eqnarray}
   V_\mu&=&\frac{1}{2}\left[\xi (v'_\mu +a'_\mu )\xi^\dagger
                      +\xi^\dagger (v'_\mu -a'_\mu )\xi\right],\nonumber \\
   A_\mu&=&\frac{1}{2}\left[\xi (v'_\mu +a'_\mu )\xi^\dagger
                      -\xi^\dagger (v'_\mu -a'_\mu )\xi\right].
\end{eqnarray}
In these variables the replacement can be written as
\begin{eqnarray}
  V_\mu&\rightarrow&V_\mu +(m-\hat{m})^2\frac{\kappa}{2}
        \left[\xi (X_\mu +Y_\mu )\xi^\dagger
             +\xi^\dagger (X_\mu -Y_\mu )\xi 
        \right], \nonumber \\
  A_\mu&\rightarrow&A_\mu +(m-\hat{m})^2\frac{\kappa}{2}
        \left[\xi (X_\mu +Y_\mu )\xi^\dagger
             -\xi^\dagger (X_\mu -Y_\mu )\xi
        \right]
\end{eqnarray}
where $X_\mu =\tau_k X_{k\mu}$, $Y_\mu =\tau_k Y_{k\mu}$ with the following
expressions for $X_{k\mu}$ and $Y_{k\mu}$
\begin{equation}
   X_{k\mu}=\varepsilon_{kji}\phi_i\partial_\mu\phi_j\frac{\sin^2\phi}{\phi^2},
\end{equation}
\begin{equation}
   Y_{k\mu}=\left[\delta_{kj}\sin\phi\cos\phi +\frac{\phi_k\phi_j}{\phi}
            \left(1-\frac{\sin\phi}{\phi}\cos\phi\right)
            \right]\frac{\partial_\mu\phi_j}{\phi}.
\end{equation}
One can obtain that
\begin{equation}
   \xi (X_\mu +Y_\mu )\xi^\dagger +\xi^\dagger (X_\mu -Y_\mu )\xi =0,
\end{equation}
\begin{equation}
   \xi (X_\mu +Y_\mu )\xi^\dagger -\xi^\dagger (X_\mu -Y_\mu )\xi 
   =2\xi_{\mu} 
\end{equation}
where
\begin{equation}
   \xi_\mu =\tau_k\xi_{k\mu}, \quad 
   \xi_{k\mu}=\left[\delta_{kj}\frac{\sin\phi}{\phi}
                   +\frac{\phi_k\phi_j}{\phi^2}
                   \left(1-\frac{\sin\phi}{\phi}\right)
              \right]\partial_\mu\phi_j .
\end{equation}
Therefore we get for the nonlinear case the standard replacement for 
the axial-vector field
\begin{equation}
   A_\mu\rightarrow A_\mu +\kappa (m-\hat{m})^2\xi_\mu .
\end{equation}
At the same time the vector field does not change.

 %%%%%%%%%%%%%%%%%% REFERENCES %%%%%%%%%%%%%%%%%%%%%%%%%%%%%%%%%%%%%%%%%%%%%%%%%
\baselineskip 12pt plus 2pt minus 2pt

\end{document}